\documentstyle[11pt,aaspp4,epsf]{article}

\def\psim{\lower.5ex\hbox{$\; \buildrel \propto \over
\sim \;$}}
\def\u{{\upsilon }}
\def\d{{\delta}}
\def\e{{\epsilon}}
\begin{document}

\title{The External Shock Model of Gamma-Ray Bursts:\\
Three Predictions and a Paradox Resolved}

\author{Charles D. Dermer\altaffilmark{1}, Markus
B\"ottcher\altaffilmark{2,1}, and James
Chiang\altaffilmark{3,1,4}}

\altaffiltext{1}{E. O. Hulburt Center for Space
Research, Code 7653,
       Naval Research Laboratory, Washington, DC
20375-5352}
\altaffiltext{2}{Department of Space Physics and
Astronomy,  Rice University, Houston, TX  77005-1892}
\altaffiltext{3}{JILA, University of Colorado,
Boulder, CO 80309-0440}
\altaffiltext{4}{NRL/NRC Resident Research Asscociate}

\begin{abstract} 

In the external shock model, gamma-ray burst (GRB) emissions are
produced by the energization and deceleration of a
thin relativistic blast wave due to its interactions
with the circumburst medium (CBM). We study the
physical properties of an analytic function which
describes temporally-evolving GRB spectra in the limit
of a smooth CBM with density
$n(x)\propto x^{-\eta}$, where $x$ is the radial
coordinate. The hard-to-soft spectral evolution and
the intensity-hardness correlation of GRB peaks are
reproduced.  We predict that (1) GRB peaks are aligned
at high photon energies and lag at low energies
according to a simple rule; that (2) temporal indices
at the leading edge of a GRB peak display a
well-defined shift with photon energy; and that (3)
the change in the spectral index values
between the leading and trailing edges of a GRB
peak decreases at higher photon energies. The
reason that GRBs are usually detected with $\nu F_\nu$
peaks in the 50 keV - several MeV range for
detectors which trigger on peak flux over a fixed time interval
is shown to be a consequence of the inverse correlation of peak 
flux and duration of the radiation emitted by decelerating blast waves. 

\end{abstract}


\section*{1. Introduction}

The {\it Beppo-SAX} results have revolutionized our
understanding of GRBs by opening a window on X-ray,
 optical, and radio
afterglows which are delayed from the prompt X-ray and
$\gamma$-ray emissions by several hours and more 
(e.g., Costa et al.\ \markcite{cea97}1997;  Feroci et
al.\ \markcite{fea98}1998; Piro et al.\
\markcite{pea98}1998; van Paradijs et al.\ \markcite{vea97}1997;
 Frail \markcite{frail98}1998).  The
decaying power-law long-wavelength afterglows are
compellingly explained as the emissions from a
relativistic blast wave which decelerates and radiates
through the process of sweeping up material from a
uniform CBM (e.g., Paczy\'nski \& Rhoads
\markcite{pr93}1993; Katz \markcite{katz94}1994; M\'esz\'aros \& Rees
\markcite{mr97}1997; Wijers, Rees, \& M\'esz\'aros
\markcite{wrm97}1997; Tavani \markcite{tavani97}1997). 
Afterglow behaviors more
complicated than simple power laws may involve effects
of extinction, self-absorption, scintillation, and CBM
structure. 

M\'esz\'aros \& Rees (\markcite{mr93}1993) proposed
that the energy of a fireball's blast wave could be
efficiently converted into radiation during the prompt
gamma-ray luminous phase of a GRB through the
interaction of a blast wave with the CBM. In the
external shock model, all GRB emissions are due to the
effects of a single thin blast wave which interacts
with the CBM. Because the directed kinetic energy of
the blast wave is converted into nonthermal particle
energy by sweeping up material from the CBM, the
emitted radiation depends crucially on the density
distribution of the CBM in the blast wave's path.
Smoother GRB time profiles therefore represent more
 uniform CBMs, at least within the Doppler cone from
which most of the detected GRB emissions originate.
Conversely, erratic and spiky GRB time profiles
represent blast-wave deceleration in highly textured
CBMs (Dermer \& Mitman
\markcite{dm99}1999). 

In a recent paper (Dermer, Chiang, \& B\"ottcher
\markcite{dcb99}1999; hereafter DCB), we proposed a
parametric description of the radiation observed from
a spherically expanding blast wave which decelerates
and is energized by sweeping up material from a smooth
CBM. By ``smooth"
we mean that the CBM density distribution is
adequately represented by the expression  $n(x) = n_0
x^{-\eta}$, where $x$ is the radial coordinate. In
this limit, the deceleration of the blast wave
produces a time profile which mimics the so-called Fast
Rise, Exponential Decay (FRED) GRB light curves. Our
{\it ansatz} is that a smooth CBM produces an ideal
FRED light curve. Conversely, a GRB which exhibits the
classic FRED-type profile results from a fireball
embedded within and expanding into a smooth CBM. 

In this {\it Letter}, we make three quantitative
predictions that can be tested using BATSE
and Beppo-SAX data from bright FRED-type GRB time
profiles, though the predictions would be most
thoroughly tested with a GRB telescope which is much
more sensitive than BATSE and measures prompt GRB
emission in the range between
$\approx 1$ keV and several MeV. In Section 2, we
describe the model equations derived in DCB. Spectral
and temporal predictions are given analytically and
described graphically for a model fireball with
$\Gamma_0 = 300$. In Section 3 we show that the reason that GRBs
are usually observed with $\nu F_\nu$ peaks in
the $\sim 50$ keV-several MeV range is understood when
account is taken of blast-wave physics and
the triggering properties of burst detectors.

\section*{2. Analytic Description and Predictions of the Blast Wave
Model}

In DBC, a parametrization of the nonthermal synchrotron
emission from a decelerating blast wave observed at time $t$
and photon energy $\e = h\nu/m_ec^2$ was proposed.  The
function
is based on the analytic approach of Dermer
\& Chiang (\markcite{dc98}1998) and the numerical
results of Chiang \& Dermer (\markcite{cd99}1999),
which were in turn based upon the blast-wave physics
developed by, e.g., Blandford \& McKee
(\markcite{bm76}1976), Rees \& M\'esz\'aros
(\markcite{rm92}1992),  Piran \& Shemi
(\markcite{ps93}1993), M\'esz\'aros \& Rees
(\markcite{mr93}1993), M\'esz\'aros, Laguna, \& Rees
(\markcite{mlr93}1993),  Waxman
(\markcite{Waxman97}1997), and Vietri
(\markcite{Vietri}1997). 

The deceleration of the relativistic blast wave is assumed to 
follow the expression $\Gamma(x) = \Gamma_0$ when $x
\leq x_d$,  and $\Gamma(x) = \Gamma_0 (x/x_d)^{-g}$ when
$x_d\leq x \leq x_d\Gamma_0^{1/g}$.  Here $\Gamma_0$ is the
initial bulk Lorentz factor of the blast wave and
the decleration radius $x_d = 2.6\times 10^{16} [(1-\eta/3) E_{54}/(n_2
\Gamma_{300}^2)]^{1/3}$ cm. In this relation, the burst source 
emits $\partial
E/\partial \Omega = 10^{54}E_{54}/(4\pi)$ ergs
sr$^{-1}$, $\Gamma_0 \equiv 300\Gamma_{300}$, and $n_2
= n_0/(10^2$ cm$^{-3})$. For simplicity, we consider
only spherically symmetric blast waves. The
parameter $g$ specifies the radiative regime, and
$g \rightarrow (3-\eta)/2$ and $g \rightarrow 3-\eta$
in the adiabatic and radiative limits, respectively. The deceleration 
time scale $t_d =  9.7 (1+z) [(1-\eta/3) E_{54}/(n_2
\Gamma_{300}^8)]^{1/3}$ s.

The function proposed in DBC to model the blast-wave radiation
 has the broken-power law form
$$P(\e,t) \; =\; 4\pi d_L^2 \nu F_\nu = \;
{(1+\u/\delta)\; P_p(t)\over
[\e/\e_p(t)]^{-\u}+(\u/\delta)[\e/\e_p(t)]^{\d} 
}\;,\eqno(1)$$
where $\u$ and $\delta$ are the $\nu F_\nu$ spectral indices at energies
 below and above the temporally-evolving
break energy $\e_p(t)$, respectively.
Expressions for  $\e_p(t)={\cal E}_0 [\Gamma(x)/\Gamma_0]^4 (x/x_d)^{-\eta/2}$
 and the time-varying amplitude 
$P_p(t)=\Pi_0 [\Gamma(x)/\Gamma_0]^4 (x/x_d)^{2-\eta}$
 are given in DBC by eqs.(9) and (16), respectively.
Eq.\ (1) depends on the
specification of the nine parameters listed in Table
1. The equipartion parameter $q = [\xi_H(r/4)]^{1/2}\xi_e^2$, where
$r$ is the compression ratio and $\xi_H$ and $\xi_e$ are the magnetic field and electron equipartition values, respectively (see eqs.\ [7] and [8] in DCB).
 The parameters $\u$, $\delta$,
$g$, $\eta$, and  $q$
are assumed to be independent of time.  This
assumption is obviously not true in general, but is
realized in the limit that high-quality broadband
measurements from a bright FRED-type GRB are made over
a sufficiently short observing times $\delta t$
 so that radiative-cooling 
and magnetic field-evolution time scales in the comoving frame
are $\gg \Gamma \delta t/(1+z)$.

The thick curves in Fig.\ 1 show time profiles
calculated at various observing energies
$\epsilon$, using the equations of the previous
section. The thin curves show spectral indices
calculated between $\epsilon$ and $2\epsilon$.   Here
we show results for fireballs with $\Gamma_0 = 300$
which are located at redshift $z = 1$; other parameters are
listed in Table 1.  In this example, $t_d = 9.6 (1+z)$
s and
${\cal E}_0 = 2.43/(1+z)$. The model GRB light curves shown in
Fig.\ 1 display a rapid rise followed by a gentler
decay which approaches a power-law afterglow behavior
$\propto t^{-1.52}$ (see eq.\ [22] in DCB) at late times
$t \gg t_d$. The overall shape of the model light curves
 shown in Fig.\ 1 resembles the characteristic
FRED-type GRB light curve (e.g., Fishman \& Meegan 
\markcite{fm95}1995). The model GRB 
 represents a very bright BATSE GRB if located
at $z = 1$, but would fall below BATSE detectablity if
located at $z \gtrsim 3$.

The analytic approximation used for the evolution of
$\Gamma(x)$ gives spectral index curves which are
constant when $t \leq t_d$. At later times, the 
blast-wave deceleration means that an observer
measuring a photon spectrum over a fixed range of
energies will sample photons which are produced at
progressively higher energies in the blast wave's
frame of reference.  Because a nonthermal synchrotron
spectrum from a power-law distribution of electrons
with a low energy cutoff is very hard at low energies
and softens at higher energies, blast-wave
deceleration produces the hard-to-soft spectral
evolution observed in many GRB time profiles (Norris et
al.\ \markcite{nea86}1986).  If the observer is
monitoring a GRB at $\epsilon \gtrsim {\cal E}_0$, the
flux is brightest when $t \approx
t_d$. As can be seen from Fig.\ 1, the spectrum
remains hard until $t \gtrsim t_d$, after which the
blast wave begins to decelerate and the spectrum
softens. Thus the spectrum is hardest when it is most
intense, accounting for the hardness-intensity
correlation observed in GRB light curves (Golenetskii
et al.\ \markcite{gea83}1983).  If the blast wave
produces short time scale variability by interacting
with inhomogeneities in the CBM (Dermer \& Mitman \markcite{dm99}1999),
 then the individual
pulses in GRB profiles would likewise exhibit spectral
hardening and subsequent softening, as generally
observed in well-defined pulses of GRB light
curves (Ford et al.\ \markcite{fea95}1995; Crider et
al.\ \markcite{cea98}1998). The qualitative ability of
the blast wave model to  explain these empirical
trends has been pointed out previously by Panaitescu \&
M\'esz\'aros (\markcite{pm98}1998). 

We now propose three quantitative
predictions to test the validity of this
model.  As is evident from Fig.\ 1, the peak
flux shifts to later times at lower photon energies.
The peak times are given
analytically by the expression
$$ t_p (\epsilon) = t_d\; \max \{ 1, (1+2g)^{-1} \left[
\left( {\epsilon \over {\cal E}_0} 
\right)^{-{2 g + 1 \over 4 g + \eta/2}} + 2 \, g
\right] \} \; \eqno(2) $$
(see also Chiang \markcite{chiang98}1998).  By
plotting the times of a well-defined peak for
a GRB measured over a large energy range, one tests
the external shock model and constrains values of
${\cal E}_0$ and the index $(2g+1)/(4g+\eta/2)$. The
peak shifting will be more pronounced in GRBs
with larger values of ${\cal E}_0$.

Our second prediction is that the temporal indices of
a pulse profile vary with photon energy according to
the relation $ \chi(\epsilon;t) =
\partial \ln P(\e,t)/ \partial \ln t $.  This yields
a complicated analytic expression from eq.\ (1),
though it is easily calculated numerically.  When 
$t < t_d$, however, the analytic form gives
$$P(\epsilon,t) = \Pi_0
\;\times  \cases{  (1+{\u \over \d }) ({\epsilon\over
{\cal E}_0 })^{-\u}(t/t_d)^{2 -
\eta +
\eta\u /2} & for $\e \ll {\cal E}_0$,~and \cr\cr 
(1+{\delta\over \u}) ({\epsilon\over {\cal E}_0
})^{-\delta}(t/t_d)^{2 - \eta -
\eta\delta/2} & for $\e \gg {\cal E}_0$. \cr}
\eqno(3)$$ Thus we predict that the temporal index
$\chi$  changes from $2-\eta(1-\u /2)$ at low energies
to $2-\eta(1+\delta/2)$  at high energies. For a
uniform CBM, there is no shift and the temporal index
$\chi = 2$. When $\eta\neq 0$, the relation
between the temporal and spectral indices indicated by
eq.\ (3) implies the value of $\eta$.

As can also be seen from Fig.\ 1, the spectral index is
hardest at the leading edge of a GRB peak and softens
at the trailing edge, with the change in the
values of the spectral index decreasing towards higher
photon energies.  This prediction can be made
quantitative by taking the derivative of eq.\ (1),
giving the result
$$\alpha (\epsilon, t)  = {\partial \ln P(\e,t)\over
\partial \ln \e }  = {\u (y^{-\u}- y^\delta ) \over
y^{-\u} + (\u/\d) y^\d }\; \eqno(4)$$ for the  $\nu
F_\nu$ spectral index, where $y = \e/\e_p(t)$. The
external shock model can be tested by plotting the
energy-dependent variation of the spectral index
across the peak of a GRB from eq.\ (4).

Although the above equations provide a simple analytic characterization
of the predictions of the external shock model, they depend crucially
on the $\Gamma(x)$-prescription noted above.
Moreover, electron cooling is not taken into account in the analytic model. 
 We used a numerical
simulation code (Chiang \& Dermer \markcite{cd99}1999) to determine the 
reliability of the analytic predictions. The results are shown in Fig. 2 using
the parameters of Table 1. Because the radiative regime $g$ in the
 simulation depends on the magnetic field $H$(G),
 the value of $\xi_H = H^2/[32\pi m_pc^2 n_0 (r/4)]$ was adjusted
 until the bulk Lorentz factor  
approached the asymptotic behavior $\Gamma\propto x^{-2}$ in the deceleration
 phase. This occurred for $\xi_H \cong 3\times 10^{-5}$.  The injection 
spectrum of electrons is chosen $\propto \gamma^{-3.4}$ in order to 
give an uncooled spectrum with $\delta = 0.2$. Electron cooling
is seen to be important at high photon energies. The peak of the light curves
at large photon energies occurs at $\approx t_d/2$, with a peak flux $\approx 2\times$ greater than the analytic estimate. 
We find that the qualitative trends of the predictions described above are 
reproduced in the detailed calculation. The analytic model can be used to 
confine the parameter range, but detailed fitting to data should
employ the more accurate numerical simulations.

\section*{3. Resolution of the
Relativistic Beaming/$\nu F_\nu$ Peak Energy Paradox}

Even if cosmic fireballs were produced
with a narrow range of values of $\Gamma_0$, their $\nu
F_\nu$ peaks  would be distributed over a wide range
of photon energies since ${\cal E}_0\propto
\Gamma_0^4$.  It has therefore been
something of a mystery to understand why GRBs are
usually detected with $\nu F_\nu$ peaks in a narrow
range of photon energies between $\sim 50$~keV and
$\sim 1$~MeV (see, e.g., Malozzi et al.\
\markcite{mea95}1995; Piran
\& Narayan \markcite{pn96}1996) without requiring
 fine-tuning of the parameters. Brainerd's
(\markcite{brainerd94}1994) Compton attenuation model,
for example, has been advanced to specifically address
this puzzle. By contrast to GRBs, the
nonthermal emissions in blazars are also thought to
originate from relativistic outflows, yet they have
$\nu F_\nu$ peaks in both their synchrotron and
Compton components which range over three
orders-of-magnitude or more in photon energy 
(e.g., von Montigny et 
al.\ \markcite{vea95}1995). 

Using the spectral form proposed in DCB, we show how
this paradox is resolved for a uniform CBM (the
generalization to $\eta \neq 0$ is straightforward).
Suppose an instrument triggers on peak flux  
measured over the time scale $\Delta t$ and over a
narrow range of photon energies centered at
$\e_d$. When the mean duration $t_p(\e_d)$ of the fireball measured
at $\e_d$ is longer than $\Delta t$, then the detector 
triggers on peak flux.  From the expressions for $\epsilon_p(t)$ and $P_p(t)$,
 we 
find that fireballs with 
$ \Gamma_0 = \bar \Gamma_0 \cong  240\;
 [ (1+z) \e_d /(n_2^{3/8} q_{-3})]^{1/4}$ 
are observed with the peak of their
$\nu F_\nu$ spectrum at photon energy $\e_d$ at the
moment when the received bolometric power is greatest
(i.e., when $t = t_d$).  Dirtier fireballs with
$\Gamma_0 < \bar \Gamma_0$ produce a peak flux
at $\e_d$ which declines according to the relation
$P_p(\e_d,t_p) \propto \Gamma_0^{(4\d +8/3)}$,
 as can be derived by inserting eq.\
(2) into eq.\ (1) in the limit $\e_d \gg {\cal E}_0$
(see also eq.\ [19] in DCB).  Because $\delta > 0$,
the peak flux therefore rapidly decreases with
decreasing $\Gamma_0$. Cleaner fireballs with $\Gamma_0
> \bar
\Gamma_0$ are also more difficult to detect because in
this case,
$P_p(\e_d,t_p) \propto \Gamma_0^{2g^{-1} - 4 /3}\;$ 
(cf. eq.\ [20] in DCB).  In order to be
observed at all, blast waves cannot be perfectly
adiabatic,  so that $g > 3/2$ and the peak flux
detected with a GRB instrument decreases with
increasing $\Gamma_0$.

Cleaner fireballs are more difficult to detect than
indicated above, because when $t_d \ll \Delta t$, the
detector triggers on fluence rather than flux. 
The fluence $F$
decreases with increasing $\Gamma_0$ according 
$F \propto t_p P_p(\e_d,t_p) \propto \Gamma_0^{g^{-1}-2/3}\times
 \Gamma_0^{2g^{-1} -4 /3}$,
using eq.\ (2) and the expression for $\epsilon_p(t)$ with $\eta = 0$. 
Because of the emission properties of blast waves,
detectors which trigger on peak flux over a fixed time window
will therefore be most sensitive to fireballs which
 have $\nu F_\nu$
peaks in the energy range of the detector. Hence the discovery of
new classes of clean and dirty fireballs must consider 
varying time
windows and energy ranges in the triggering criteria of a 
detector, 
and must also contend with the levels of background radiation as
discussed in DCB.

In separate work (B\"ottcher \& Dermer \markcite{bd99}1999), we model the 
triggering properties of GRB detectors and show that compared to the 
BATSE GRB trigger rate, 
 the dirty fireball rate is poorly known due to selection biases 
against their detection. Observational analyses of {\it Ariel V} X-ray data 
(Grindlay \markcite{g99}1999; see discussion in DCB)
 provides the strongest available 
limit on the frequency of dirty fireballs. By contrast, the relative 
rate of clean fireballs is strongly constrained in the analysis of 
B\"ottcher \& Dermer 
(\markcite{bd99}1999) unless there is a strong anti-correlation between
 $\Gamma_0$ and $q$.

In summary, we have used a simple analytic description
of the external shock model of GRBs to pose three
predictions which can be tested by fitting eqs.\
(2)-(4) 
to high quality data obtained with a GRB
telescope sensitive in the $\approx 1$ keV -
several MeV range.  Although we have employed a simple
mathematical parametrization, detailed fits to data
should use numerical simulations of the
decelerating blast wave (e.g., Chiang \& Dermer
\markcite{cd99}1999, Panaitescu \& Meszaros
\markcite{pm98}1998).  We have also explained why
detectors which trigger on peak flux are most
sensitive to fireballs which produce $\nu F_\nu$ peaks
in the 50 keV - several MeV range, resolving a
long-standing puzzle in relativistic beaming models of
GRBs.

\acknowledgments
We thank the referee for comments. 
The work of CD and MB was partially supported by
the {\it Compton Gamma Ray Observatory} Guest 
Investigator Program. The work of JC was performed while he
held a National Research Council - Naval Research Laboratory
Associateship.  CD acknowledges support of the Office
of Naval Research.

\begin{table}[h]
\small
\caption{Standard Parameters for Model of Evolving GRB
Spectra}
\bigskip
\begin{tabular}{c l l c}
\tableline\tableline Parameter  & Standard Value &
Description \\
  &  \\
\tableline
$\Gamma_0$ & 300 & baryon-loading
parameter \\
$q$ & $10^{-3}$ & equipartition term \\
$g$ & 2 & index of $\Gamma$ evolution \\
$\partial E_0/\partial \Omega$  &
$10^{54}E_{54}/(4\pi)$ & total fireball energy
released per unit solid angle, ergs sr$^{-1}$ \\
$\u$ & 4/3 & spectral index of rising portion of $\nu
L_\nu$ spectrum \\
$\delta$ & 0.2 & spectral index of falling portion of
$\nu L_\nu$ spectrum \\
$z$ & 1 & cosmological redshift \\
$n_0$  & $10^2$ cm$^{-3}$ & density at $ x_d$ \\
$\eta$  &  0 & index of density distribution\\
\tableline
\end{tabular}
\label{table1}
\end{table}

\eject

\setcounter{figure}{0}
\begin{figure}
\epsfysize=12cm
\epsffile[0 20 500 500]{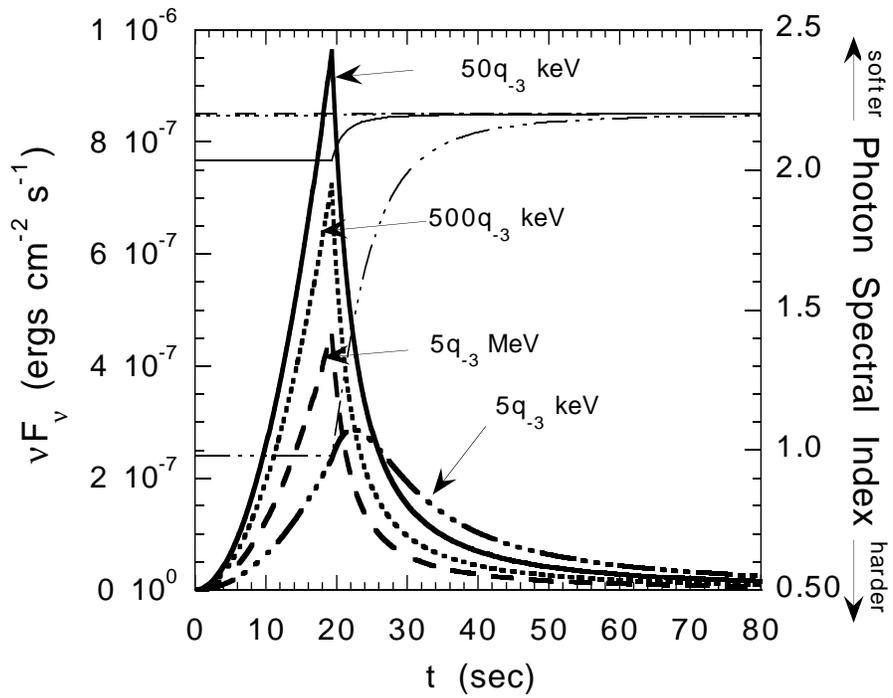}
\caption[]{Light curves (thick lines) and spectral indices (thin lines) from a
model GRB produced by a fireball with $\Gamma_0$ =
300 located at redshift $z = 1$.  Other parameters are
listed in Table 1. }
\end{figure}

\setcounter{figure}{1}
\begin{figure}
\epsfysize=12cm
\epsffile[0 20 500 500]{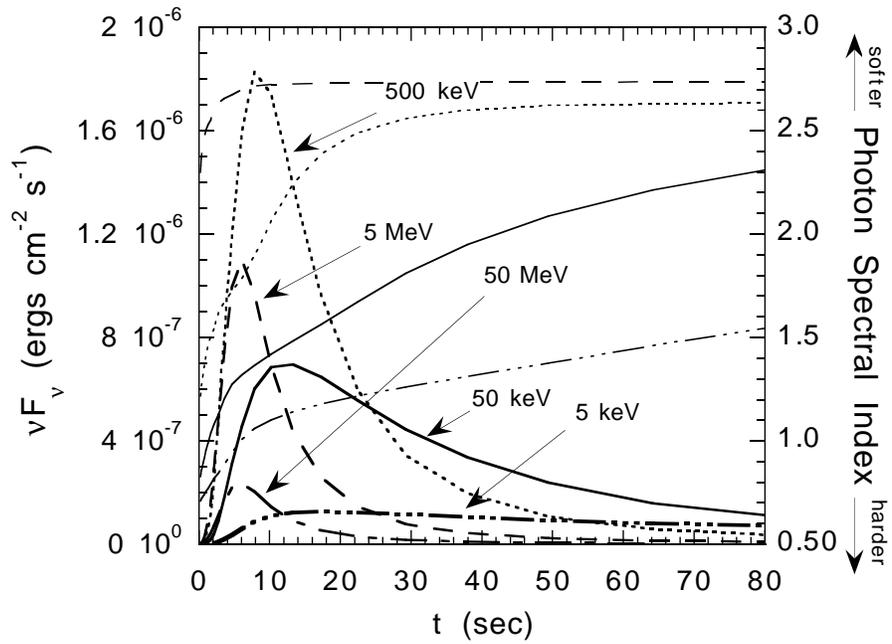}
\caption[]{Numerical simulation results corresponding to the
 analytic model shown in Fig. 1 (see text for details).
 Equipartition between the energy in electrons, protons, and magnetic field is assumed. }
\end{figure}

\end{document}